\documentclass[letterpaper,english,reprint, aps]{revtex4-1}
\pdfoutput=1
\usepackage[T1]{fontenc}
\usepackage[latin9]{inputenc}
\usepackage{amsmath}
\usepackage{amssymb}
\usepackage{graphicx}
\usepackage{esint}
\usepackage{cleveref}

\makeatletter

\pdfpageheight\paperheight
\pdfpagewidth\paperwidth

 \@ifundefined{textcolor}{}
 {%
   \definecolor{BLACK}{gray}{0}
   \definecolor{WHITE}{gray}{1}
   \definecolor{RED}{rgb}{1,0,0}
   \definecolor{GREEN}{rgb}{0,1,0}
   \definecolor{BLUE}{rgb}{0,0,1}
   \definecolor{CYAN}{cmyk}{1,0,0,0}
   \definecolor{MAGENTA}{cmyk}{0,1,0,0}
   \definecolor{YELLOW}{cmyk}{0,0,1,0}
 }

\makeatother

\usepackage{babel}
\begin{document}

\title{Bandwidth Allocation and Resource Adjustment for Stability Enhancement in Complex Networks}

\author{K. Y. Henry Tsang}

\affiliation{Department of Physics, The Hong Kong University of
  Science and Technology, Clear Water Bay, Kowloon, Hong Kong}

\author{K.Y. Michael Wong}

\affiliation{Department of Physics, The Hong Kong University of
  Science and Technology, Clear Water Bay, Kowloon, Hong Kong}

\date{\today}
\begin{abstract}
We introduce the discrete Green's function to elucidate how resource fluctuations determine flow fluctuations in a network optimizing a global cost function. To enhance the robustness of the network against fluctuations, we develop the schemes of optimal bandwidth allocation in links and optimal resource adjustment in nodes. With the total bandwidth of the network fixed, the approach of optimal bandwidth allocation is to increase the bandwidth in links such that the number of overloaded links or the amount of excess flows in networks under fluctuations can be minimized. Similarly, the approach of optimal resource adjustment is to minimize the number of overloaded links in networks under fluctuations with the total resource change in the network fixed. Compared with the conventional approach of proportionate bandwidth assignment or resource reduction, it is found that the optimized bandwidth allocation or resource adjustment can highly enhance the stability of the networks against fluctuations. The changes of loads and currents prescribed by the optimal bandwidth allocation and resource adjustment schemes are correlated with each other, except for some nodes that exhibit relay effects.  
\end{abstract}
\maketitle


\section{\label{sec:Inroduction}Introduction}

Stability of power grids is essential for the development of modern societies. Due to the advancement of technology over years, power grids become more reliable and stable in supplying electricity. However, widespread blackouts of different scales in power grids still occurred frequently \citep{carreras2016north} even with the investment of advanced technology. In spite of the fact that blackouts in large power grids are rare, they cause overwhelming economic and social losses \citep{2econ}. As a result, it is crucial to study how the robustness of the networks can be enhanced. There are also new challenges about the stability in power grid systems in recent years due to the introduction of renewable energy sources, such as wind power and solar power, and the growing trend of power trading. Renewable energy usually has strong fluctuations, and such increasing deployment of renewable energy in power grids endanger the stability of the networks \citep{schafer2018non}. This is because power grids usually operate near their capacity limits. Flow fluctuations can cause the current flow to exceed the capacity of the transmission cables. Seeing that a small number of link failures can cause a large failure in the network, it can largely affect the stability of the network and it becomes important to study the flow fluctuations in the network \citep{16fluct_induce}. Hence it is important to incorporate fluctuations in the formulation of the optimal power flow problem \citep{bienstock2014chance}. In this work, we focus on a typical model of resource allocation that can be formulated as a power grid network in the direct current approximation and study techniques to strengthen the stability of the network against fluctuations. This is useful for the design and control of the power grid, in which future estimates of resources with uncertainty are predicted.

Many networks used for resource transportation or communication have some capacity (or bandwidth, depending on the type of the network) assigned to a node (a link) to withstand the flows through it. One common way to increase the robustness of the network is to increase the capacities of nodes or bandwidths of links. A plausible suggestion was to allocate the additional bandwidth proportional to the initial loads \citep{motter2002cascade}. However, as we shall see, the proportionate increase in capacities is not effective and costly to enhance the stability of the networks. Therefore, in this work, we study improved schemes of allocating bandwidths in the links of the network to prevent cascading failures caused by fluctuations. 

In general, allocating bandwidths is more useful in the network designing stage. For real-time control, it is more practical to maneuver the resources in the network. To relieve the stress in the network, demand has to be shed from the network and the power generators have to be ramped down. This is often known as load shedding in power engineering \citep{bienstock2011optimal}. Ideally, the population of affected consumers should be minimized at the same time and the objective is then to minimize the overload probabilities of cables while the load volume to be shed are bounded by a chosen threshold. For such a purpose, we also study an optimal scheme of adjusting the resource in the network to enhance the stability of the network.

In Sec. II, we introduce the model for the networks and discrete Green's function for calculating the network flow. The discrete Green's function are then used for predicting the variance of flow fluctuations given the information of resource fluctuations. In Sec. III, using the predicted variance of the flow fluctuations, we develop the optimized bandwidth allocation against fluctuations to increase the stability of the network. In Sec. IV, we formulate an optimal resource adjustment scheme in the form of a constrained optimization problem to enhance the robustness of the network against fluctuations. The summary is presented in the final section.


\section{\label{sec:The-Min-Cost-Network-Flow-Model}The Min-Cost Network Flow Model}

We begin with a typical resource allocation network model similar to \citep{23sparse_graph}. Consider a network with $N$ nodes and each node $i$ is assigned with resource $\Lambda_i$. Positive and negative values of $\Lambda_i$ indicate that node $i$ has a supply and demand of resources respectively. Nodes with $\Lambda_i =0$ relay the current flows only. For simplicity, the network does not have excess resources (i.e. $\sum_i \Lambda_i =0$). Let $y_{ij} \equiv -y_{ji}$ be the current flow from node $j$ to node $i$ and it has to satisfy the conservation of flows,
\begin{equation}
\Lambda_i + \sum_{j \in \partial i} y_{ij} =0 , \label{conc}
\end{equation}
where $\partial i$ is the set of neighbors of node $i$. Besides satisfying the flow conservation constraint, the current flows are required to minimize the total transportation cost in which the transportation cost along link $(ij)$ is given by a general even function $\phi(y_{ij})$. In general, cost functions of form $\phi(y_{ij})= y_{ij}^r$ can be convex or concave functions of $y_{ij}$ depending on the modeling purpose. To avoid congestion in network, $ r > 1$ is usually used as it will penalize overlaps of flows while $r < 1$ is used when we want to encourage overlaps of flows \cite{bohn2007structure, yeung2013physics}. To find the optimal current flows that minimize $E=\sum_{(ij)}\phi(y_{ij})$ subject to the constraint Eq. \eqref{conc}, we introduce an updating method, the chemical potential method, to solve the flow optimization problem.

\subsection{Chemical Potential}

In the chemical potential method, we calculate the optimal flows by computing the chemical potentials for each nodes. The Lagrangian for optimization is given by
\begin{equation}
L = \sum_{(ij)} \phi(y_{ij}) + \sum_i \mu_i \left(\Lambda_i + \sum_{j \in \partial i} y_{ij} \right),
\end{equation}
where $\mu_i$ is the Lagrange multiplier ensuring the conservation of resources at node $i$. Following the convention in thermodynamics that interprets $\mu_i$ as the conjugate variable of resources, $\mu_i$ is referred to as the chemical potential of node $i$. Optimizing $L$ with respect to $y_{ij}$, we obtains the current flows as
\begin{equation}
\label{general_flow}
y_{ij} = [\phi']^{-1}(\mu_j - \mu_i),
\end{equation}
where $\phi'$ is the derivative of $\phi$ with respect to its argument. The chemical potential $\mu_i$ is given by the zero of the equation given by
\begin{equation}
\label{conser_full}
g_i(x) = \Lambda_i + \sum_{j \in \partial i} [\phi']^{-1}(\mu_j - x).
\end{equation}

In this work, we focus on the transportation cost as a quadratic function and the total cost function is given by
\begin{equation}
\label{totale}
E = \sum_{(ij)} \frac{y_{ij}^2}{2} .
\end{equation}
The quadratic cost function is chosen as it is minimized in electrical networks according to Thomson's principle \cite{doylerandom} and it will be shown to be equivalent to power grid networks in the direct current (DC) approximation. For the quadratic cost function, using Eq. \eqref{conser_full}, one can obtain the chemical potentials as
\begin{equation}
\label{laplacian}
\sum_j L_{ij}\mu_j = \Lambda_i,
\end{equation}
where $L$ is the Laplacian matrix given by
\begin{equation}
L = D - A.
\end{equation}
In the above equation, $D$ is the $N \times N$ diagonal matrix with diagonal elements $D_{ii}$ being the degree $d_i$ of node $i$, and $A$ is the adjacency matrix. In matrix notation, Eq. \eqref{laplacian} is equivalent to
\begin{equation}
\label{chem_matrix}
L \mu = \Lambda,
\end{equation}
where $\mu$ and $\Lambda$ are the column vectors of the chemical potentials and resources respectively. The chemical potentials are obtained by inverting Eq. \eqref{chem_matrix},
\begin{equation}
\mu = G \Lambda,
\end{equation}
where $G$ is the pseudoinverse of $L$, which is also known as the discrete Green's function \cite{chung2000discrete}. Hence, from Eq. \eqref{general_flow}, the current flowalong a link from node $j$ to $i$ is given by
\begin{equation}
\label{current_chem}
y_{ij} = \sum_l (G_{jl} - G_{il})\Lambda_l.
\end{equation}
Although the discrete Green's functions are nonlocal, they can be calculated once during the initial stage and subsequent estimations can be calculated by linear products. This greatly simplifies the centralized algorithm. In fact, the discrete Green's function is equivalent to the widely used Power Transfer Distribution Factors (PTDF) matrix for linear power flows \citep{ronellenfitsch2016dual}. The chemical potentials of the nodes are obtained by its multiplication with the column vector of power injection. For unit line susceptances in the network, a further multiplication by the incidence matrix maps the difference in the chemical potentials of the link vertices to the power flows of the links in the DC approximation (the incidence matrix is written out explicitly in Eq. \eqref{current_chem}).

\subsection{Power Grids}
The above techniques can be applied to study power grids in the DC approximation. To see this, we consider the power flow along a link from node $j$ to $i$ given by a sine function of the phase angle difference between two nodes \citep{wood2012power} 
\begin{equation}
P_{ij} = \frac{|V_i||V_j|\text{sin}(\theta_j -\theta_i)}{x_{ij}},
\end{equation}
where $x_{ij}$ is the reactance of the transmission link $(ij)$, $\theta_i$ is the phase angle and $V_i$ is the voltage for node $i$. In the DC approximation, differences between the phase angles for each pair of neighboring nodes are small and we can approximate the power flow equation as 
\begin{equation}
\label{phased}
P_{ij} \approx \frac{\theta_j -\theta_i}{x_{ij}},
\end{equation}
where the voltage for each node is conventionally chosen to be $|V_i| \approx 1$ with a suitable unit. Denote $P_i$ as the power generation (or consumption) in node $i$, then the total power flow in node $i$ is given by
\begin{equation}
P_i + \sum_{j \in i} \frac{\theta_j -\theta_i}{x_{ij}} = 0.
\end{equation}
From Eq. \eqref{conser_full} and Eq. \eqref{phased}, we can view the chemical potentials as the phase angles for the nodes and having a unit reactance for each transmission link. Furthermore, the flow conservation equation is equivalent to the power flow equation in a node in which the power generation or consumption can be treated as the node's resource. Thus, networks using the quadratic cost function together with the flow conservation constraint can be treated as power grid networks in the DC approximation.

\subsection{Induced Flow Fluctuations} 
In this section, we study the flow fluctuations induced by the fluctuations in resources. To begin with, we write the resource of node $i$ as the sum of two components,  
\begin{equation}
\label{fluc1}
\Lambda_i =\Lambda_i^0 +\delta\Lambda_i,
\end{equation}
where $\Lambda_i^0$ is the the original resource without fluctuation and  $\delta \Lambda_i$ is the resource fluctuations. The current flows $y_{ij}^0$ are the original current flows without fluctuation dependent on $\Lambda_i^0$ and $\delta y_{ij}$ are the flow fluctuations dependent on $\delta\Lambda_i$. For small fluctuations we can calculate the mean and variance of the flow fluctuations by using the discrete Green's function. For simplicity, we assume that the resource fluctuations are independent and the mean of resource fluctuations is zero $\left\langle \delta \Lambda_i \right\rangle = 0$. Moreover, $\left\langle \delta\Lambda_i^2 \right\rangle$ of the resource fluctuations are assumed to be known and used in estimating the flow fluctuations. 

From Eq. \eqref{current_chem}, the current fluctuations in link $(ij)$ can be obtained by 
\begin{equation}
\label{delta_ychem}
\delta y_{ij} = \sum_{l} (G_{jl}-G_{il}) \delta\Lambda_l.
\end{equation} 
The mean $\left\langle \delta y_{ij} \right\rangle$ is obtained by taking average of Eq. \eqref{delta_ychem} and is equal to 0 as $\left\langle \delta \Lambda_i \right\rangle = 0$. The second moment $\left\langle (\delta y_{ij})^2 \right\rangle$ is obtained by taking the squares and then averaging on both sides of Eq. \eqref{delta_ychem} 
\begin{equation}
\label{var_ychemsim}
\langle {\delta 
y_{ij}}^2 \rangle=\sum_{k} {(G_{jk} -G_{ik})}^2 \langle {\delta 
\Lambda_k}^2 \rangle .
\end{equation}


\section{\label{sec:Optimal-Bandwidth-Allocation}Optimal Bandwidth Allocation}

\subsection{Proportionate Bandwidth Allocation}

Estimates of flow fluctuations are useful in allocating bandwidths to links in networks handling fluctuating traffic. In practice, each link has a capacity which allows the maximum amount of current before it becomes overloaded. In communication networks, such capacities are usually referred to as bandwidths and we generalize such terminology in this paper for convenience. Usually the bandwidth of a link is designed to be able to withstand the current flow of the network in the steady state with some tolerance. The reason for the tolerance is to prevent overload by a sudden increase or fluctuations in current flow. For simplicity, we decompose the bandwidth of a link into two parts,
\begin{equation}
L_{ij} = L_{ij}^0 + \Delta L_{ij},
\end{equation}
where $L_{ij}^0 = |y_{ij}^0|$ and $\Delta L_{ij} \geq 0$ represents the tolerance. A plausible bandwidth allocation scheme is the proportionate bandwidth which has the form
\begin{equation}
\label{og_band}
L_{ij} = (1+\varepsilon)L_{ij}^0,
\end{equation}
where $\varepsilon$ is the tolerance factor and $\Delta L_{ij} = \varepsilon L_{ij}^0$. The proportionate bandwidth has been adopted to study the robustness of networks due to its simple proportionate form \citep{motter2002cascade}. In practice, increasing the bandwidths will increase the cost of constructing the network and hence, $\varepsilon$ is usually desired to be small.
Therefore, for the purpose of designing a stable network, one is interested in increasing the robustness of the network by restricting $\Delta L_{ij}$ with a given average tolerance factor (i.e. $\sum_{(ij)} \Delta L_{ij} = \varepsilon \sum_{(ij)} L_{ij}^0$). For a given tolerance factor, nevertheless, allocating bandwidth using the form of Eq. \eqref{og_band} is not effective against flow fluctuations.

To illustrate this, simulations in different network structures with the proportionate bandwidth allocation are used for investigation. In the simulations and analysis, we specify that the resource fluctuations at a node is unbiased and has a variance proportional to the square of the resource, that is, $\langle \delta\Lambda_i \rangle = 0$ and $\langle \delta\Lambda_i^2 \rangle = k^2 (\Lambda_i^0)^2$. We will refer to $k$ as the fluctuation parameter. When there are fluctuations in the current flows, the current may exceed the bandwidth and the link is considered overloaded when $|y_{ij}| > L_{ij}$. The fraction of overloaded links is used as a measurement of the robustness of the networks against fluctuations. We perform simulations in the Erd\"os-Renyi (ER) network \citep{latora2017complex}, Barab\'asi-Albert (BA) network \citep{latora2017complex}, IEEE Reliability Test System 96 (RTS 96) power network \citep{grigg1999ieee} and the IEEE 300-Bus System \citep{ieee300}. The ER network is constructed by $N=1000$ nodes with the connecting probability $p=0.005$ and the BA network with $N=1000$ is constructed by attaching two edges from a new node to existing nodes with the probability $p=d_i/ \sum_j d_j$ that a new node will be connected to node $i$. The resource fluctuations in the simulations are introduced at the same instant of time to each sample having the same network structure.

\begin{figure}
\begin{centering}
\includegraphics[scale=0.45]{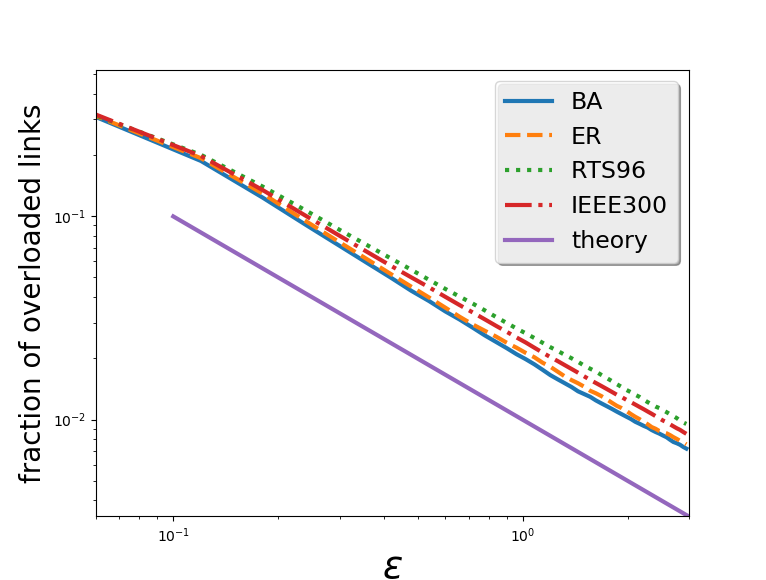}
\par\end{centering}
\caption{\label{fig:og_band}Simulation results of the fraction of overloaded links in the proportionate bandwidth allocation scheme as a function of the tolerance factor $\varepsilon$ in the power law scale for ER network with $N=1000$, BA network with $N=1000$, the RTS 96 power network and the IEEE 300-Bus System. $k=0.1$. Purple line: Analytical result. The analytical result is displaced downwards after divided by 2 for visualization.}
\end{figure}

From Fig. \ref{fig:og_band}, it can be seen that the fraction of overloaded links scales as a power law of $\varepsilon$ with exponent $-1$. This existence of non-vanishing long tails shows that the proportionate bandwidth allocation is not effective against fluctuations. The main reason is the existence of links with small average current flow, yielding only small bandwidth increase even for large values of $\varepsilon$. Furthermore, it can be seen that the existence of non-vanishing long tails occurs in various network structures. As a result, using proportionate bandwidth allocation is not efficient against flow fluctuations.

Assuming that the link fluctuations are independent, we derive analytical results of the fraction of overloaded links in Appendix. As shown in Fig. \ref{fig:og_band}, the analytical result has an excellent agreement with simulations. The fraction of overloaded links scales as the power law $\varepsilon^{-1}$ in the limit of large $\varepsilon$, but the convergence is already evident for moderate values of $\varepsilon$.

\subsection{Minimizing the Fraction of Overloaded Links}

To develop a better allocation of bandwidths, we formulate an optimization problem subject to the constraint of fixed additional bandwidths. In this work, we focus on two objective functions: the expected number of overloaded links and the total amount of excess current in the network.

\begin{figure}
\begin{centering}
\includegraphics[scale=0.5]{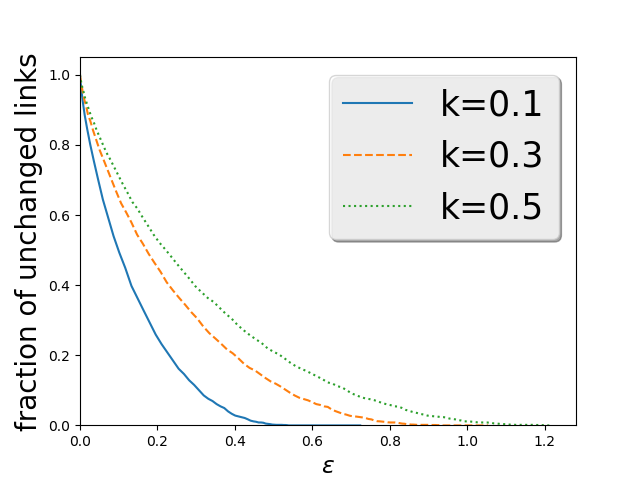}
\par\end{centering}
\caption{\label{fig:undeploy}Dependence of the fraction of unchanged links on the tolerance factor $\varepsilon$ for different fluctuation parameters.}
\end{figure}

The functional form representing the expected number of overloaded links is determined by the probability distribution of the flow fluctuations. Considering $\delta y_{ij}$ in the direction of increasing $y_{ij}^0$, it is sufficient to consider the fraction of distribution with $\delta y_{ij} > \Delta L_{ij}$ for distributions $P_{ij}(\delta y_{ij})$ of the fluctuations. Thus, the expected number of overloaded links is given by  
\begin{equation}
\label{num_fail}
F = \sum_{(ij)} \int_{\Delta L_{ij}}^{\infty} P_{ij}(\delta y) d(\delta y). 
\end{equation} 
$\Delta L_{ij}$ can then be evaluated such that it can minimize Eq. \eqref{num_fail} subject to the constraints
\begin{equation}
\label{constraint_link}
\sum_{(ij)} \Delta L_{ij} = \varepsilon \sum_{(ij)} L_{ij}^0 \text{  and } \Delta L_{ij} \geq 0.
\end{equation} 
The Lagrangian for optimization is given by 
\begin{align*}
\label{Lagrange_link}
L &= \sum_{(ij)} \int_{\Delta L_{ij}}^{\infty} P_{ij}(\delta y)d(\delta y) - \lambda \left( \sum_{(ij)} \Delta L_{ij} - \varepsilon \sum_{(ij)} L_{ij}^0 \right) \\
&+ \sum_{(ij)} \gamma_{ij} \Delta L_{ij},
\end{align*}
where $\lambda$ is the Lagrange multiplier and $\gamma_{ij}$ is the K\"uhn-Tucker multiplier. Differentiating with respect to $\Delta L_{ij}$, one can obtain
\begin{equation}
\label{link_band_general}
P_{ij}(\Delta L_{ij})=\lambda - \gamma_{ij} \text{  or  } \gamma_{ij}\Delta L_{ij} =0.
\end{equation}
Assuming that $P_{ij}(\delta y_{ij})$ are Gaussian distributions, $\Delta L_{ij}$ that can minimize the total number of overloaded links is given by 
\begin{equation} 
\label{link_band_cond}
\Delta L_{ij} =
\begin{cases}
\sqrt{\left\langle \delta y_{ij}^2 \right\rangle} \left[ 2 \ln \left( \frac{1}{\lambda\sqrt{2\pi \left\langle \delta y_{ij}^2 \right\rangle }} \right) \right]^\frac{1}{2} ,   &\text{if }\langle\delta y_{ij}^2\rangle \le \frac{1}{2\pi\lambda^2},\\
0 , &\text{otherwise}.
\end{cases}
\end{equation}
where $\lambda$ is determined by the constraint Eq. \eqref{constraint_link}. The conditional form of Eq. \eqref{link_band_cond} is due to K\"uhn-Tucker condition. For small given total cost, resources are allocated to those links with small flow fluctuations, $\sqrt{\left\langle\delta y_{ij}^2 \right\rangle} < 1 / \sqrt{2\pi}\lambda $ (unlike the proportionate bandwidth allocation scheme which allocates the bandwidth resources to every links independent of the magnitude of flow fluctuations). When the tolerance factor $\varepsilon$ is increased, the non-linearity of Eq. \eqref{link_band_cond} tends to distribute more bandwidth resources to links with moderate flow fluctuations to save the majority of links. Figure \ref{fig:undeploy} illustrates the fraction of unchanged links ($\Delta L_{ij} =0$) as a function of the total cost in a random network with fixed connectivity $d_i=3$. It shows that when the fluctuation parameter $k$ increases, the network requires higher tolerance to reduce the number of unchanged links. 

We compare the performances of the bandwidth allocation schemes by simulations in the RTS 96 network. For simplicity, {\it proport}, {\it minlink} and {\it minflow} (see next subsection) respectively correspond to the proportionate bandwidth allocation scheme, the scheme that minimizes the expected number of overloaded links, and the scheme that minimizes the expected amount of excess flows. In the simulations, the resource $\Lambda_i^0$ for node $i$ is picked from the Gaussian distribution with mean 0 and standard deviation 10, minus the average to ensure $\sum_i \Lambda_i^0 = 0$. As shown in Fig. \ref{fig:ER_compare}(a), {\it minlink} has the least fraction of overloaded links. Equivalently, {\it minlink} is able to completely protect the network with a smaller amount of total cost when compared with the proportionate bandwidth allocation scheme. As a reference, we recall that for the proportionate bandwidth allocation scheme in Fig. \ref{fig:og_band}, the total fraction of overloaded links in the network do not converge to zero even up to $\varepsilon \sim 3$.

We also consider the analytical results of the allocation of bandwidths in Appendix. For Gaussian resource fluctuations, the currents have Gaussian distributions. As shown in Fig. \ref{fig:ER_compare}(a), their agreement with the simulation results is excellent.

In the limit of large $\varepsilon$, the $\varepsilon^{-1}$ behavior of the proportionate allocation scheme is much slower than the Gaussian tail convergence in the {\it minlink} scheme, in which the fraction of overloaded links effectively scales as $\exp({-\varepsilon^2/2k^2})$. As the derivation of {\it minlink} does not require assumptions on the particular structure or size of the network, {\it minlink} can work well for all network structures (and sizes).

However, the criterion of minimizing the expected number of overloaded links saves the additional bandwidths by focusing on the links with small fluctuations, whereas it allows those links with large fluctuations to be heavily overloaded. In practice, this may lead to a huge degradation of the quality of network service. Therefore, alternative cost functions should be considered.

\subsection{Minimizing the Excess Current}

An alternative cost function for the robustness of the network is the expected amount of excess current flows. Excess current flow is defined as the amount of current flow exceeding the bandwidth 
\begin{equation}
\label{excess_y}
I_{ij} = \max(\delta y_{ij} - \Delta L_{ij} ,0).
\end{equation}
Hence, the function measuring the total excess current flow in the network can be obtained as
\begin{equation}
\label{total_excess}
F = \sum_{(ij)} \int_{\Delta L_{ij}}^{\infty} (\delta y - \Delta L_{ij}) P_{ij}(\delta y) d(\delta y) 
\end{equation}
Minimizing Eq. \eqref{total_excess} subject to the total cost constraint as Eq. \eqref{constraint_link}, one can obtain 
\begin{equation}
\label{band_flow_general}
\int_{\Delta L_{ij}}^{\infty} P_{ij}(\delta y)d(\delta y)=\lambda - \gamma_{ij} \text{  or  } \gamma_{ij}\Delta L_{ij} =0.
\end{equation}
If the flow fluctuations follow the Gaussian distributions, $\Delta L_{ij}$ that can minimize the total excess current flows is given by 
\begin{equation}
\Delta L_{ij} = \varepsilon'\sqrt{\left\langle (\delta y_{ij})^2 \right\rangle}
\label{band_flow}
\end{equation}
where $\varepsilon'$ is a function of the Lagrange multiplier and can be calculated by substituting $\Delta L_{ij} $ into Eq. \eqref{constraint_link}. Unlike the bandwidth allocation scheme that minimizes the expected number of overload links, Eq. \eqref{band_flow} tends to distribute the bandwidth resources according to the flow fluctuations in the links. In fact, the form of Eq. \eqref{band_flow} is similar to Eq. \eqref{og_band} where $L_{ij}^0$ is replaced by $\sqrt{\left\langle (\delta y_{ij})^2 \right\rangle}$. We refer to this optimization scheme as {\it minflow}.

\begin{figure}
\begin{centering}
\includegraphics[scale=0.45]{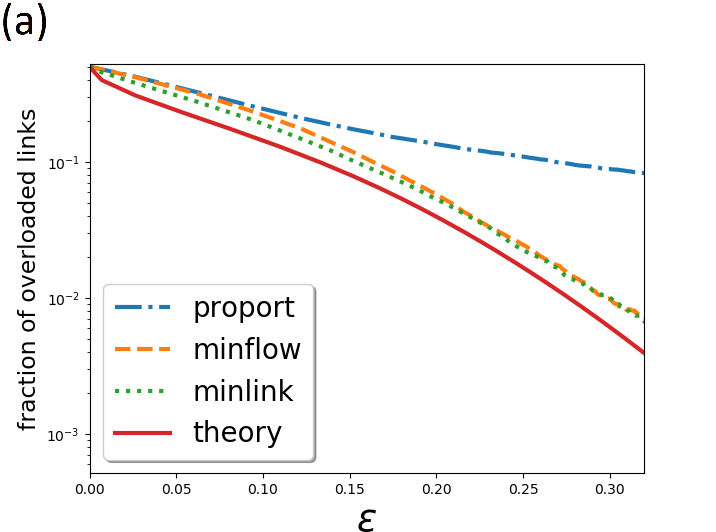}
\includegraphics[scale=0.45]{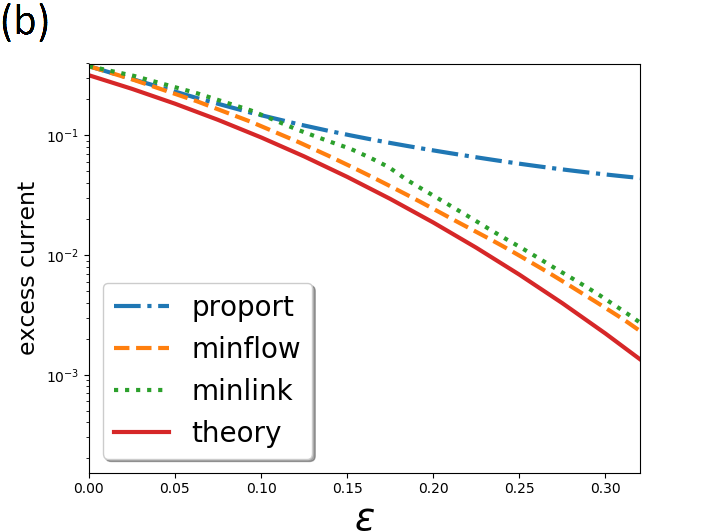}
\par\end{centering}
\caption{\label{fig:ER_compare}Simulation results of (a) the fraction of overloaded links and (b) the amount of excess current per link, in units of $\sqrt{\langle \delta y_{ij}^2 \rangle}$, for the bandwidth allocation schemes in the RTS96 network as a function of the tolerance factor $\varepsilon$. $k = 0.1$ and 100 samples of supply and demand are generated from the resource distribution.}
\end{figure}

As shown in Fig. \ref{fig:ER_compare}(b) for the RTS 96 network, the excess current in {\it minflow} has the lowest expected excess current when compared with {\it proport}. Equivalently, {\it minflow} is able to completely protect the network with a smaller amount of total cost. The simulation results also have an excellent agreement with the theoretical results.

As derived in Appendix in the limit of large $\varepsilon$, the excess current in {\it minflow}  effectively scales as $\exp({-\varepsilon^2/2k^2})$, converging much faster than the $\varepsilon^{-1}$ behaviour of the proportionate allocation scheme.

\subsection{Non-Gaussian Fluctuations}

So far we only consider the resource fluctuations following Gaussian distributions. For flow fluctuations following non-Gaussian fluctuations, one can obtain the optimal bandwidth allocation scheme by modifying the distribution in Eq. \eqref{link_band_general} and Eq. \eqref{band_flow_general}. It was recently found that the power grid frequency fluctuations are better described as L\'evy-stable distribution than Gaussian distribution \citep{schafer2018non}. Therefore, we further study the resource fluctuations following the L\'evy-stable distributions, which take the form
\begin{equation}
P(\delta\Lambda|c) = \int_{-\infty}^\infty \frac{dk}{2\pi} e^{-ik\delta\Lambda}\varphi(k|c),
\end{equation}
where $\varphi(k|c)$ is given by
\begin{equation}
	\varphi(k|c) = e^{-(c|k|)^\alpha}
\end{equation}
and $c>0$ is the scale parameter of the distribution. The case with the stability parameter $\alpha = 2$ corresponds to the Gaussian distribution, but real power grid data shows that $\alpha$ deviates from 2, Following real data we consider the case $\alpha = 1.898$. 

We note that the L\'evy-stable distribution satisfies the closure property, that is, for $x_1\sim P(x_1|c_1)$ and $x_2\sim P(x_2|c_2)$, the sum $x_1+x_2$ also obeys the L\'evy-stable distribution with the same stability parameter and with the scale parameter $c$ given by
\begin{equation}
	c=(c_1^\alpha + c_2^\alpha)^{1/\alpha}.
\end{equation}
Hence from Eq. (10), the flow fluctuations $\delta y_{ij}$ also follow the L\'evy-stable distributions with the same stability parameter and the scale parameter given by
\begin{equation}
	c_{ij} = (\sum_l(G_{jl} - G_{il})^\alpha)^{1/\alpha}.
\end{equation}
The {\it minlink} algorithm is then given by
\begin{equation} 
\Delta L_{ij} =
\begin{cases}
P^{-1}(\lambda|c_{ij}) ,   &\text{if     } c_{ij}\le \frac{\Gamma(1/\alpha)}{\pi\alpha\lambda},\\
0 ,  &\text{otherwise}.
\end{cases}
\label{eq:Levy_minlink}
\end{equation}
where $P^{-1}(\lambda|c)$ is the inverse function of $P(x|c)$.

\begin{figure}
\begin{centering}
\includegraphics[scale=0.45]{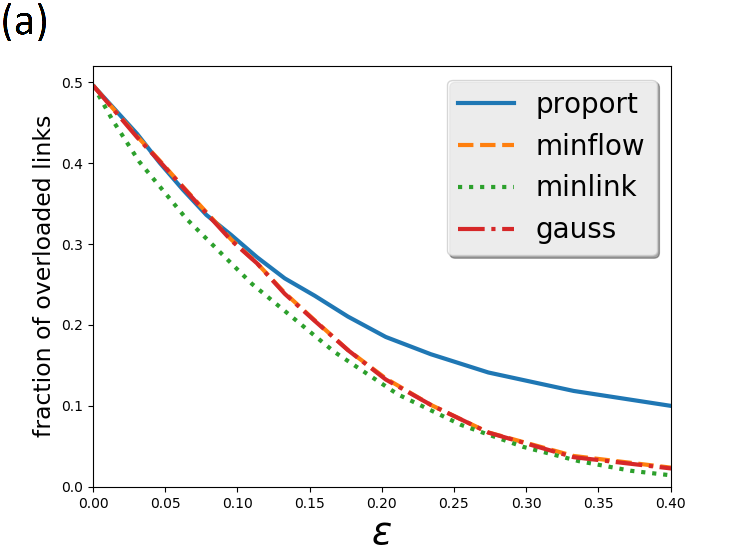}
\includegraphics[scale=0.45]{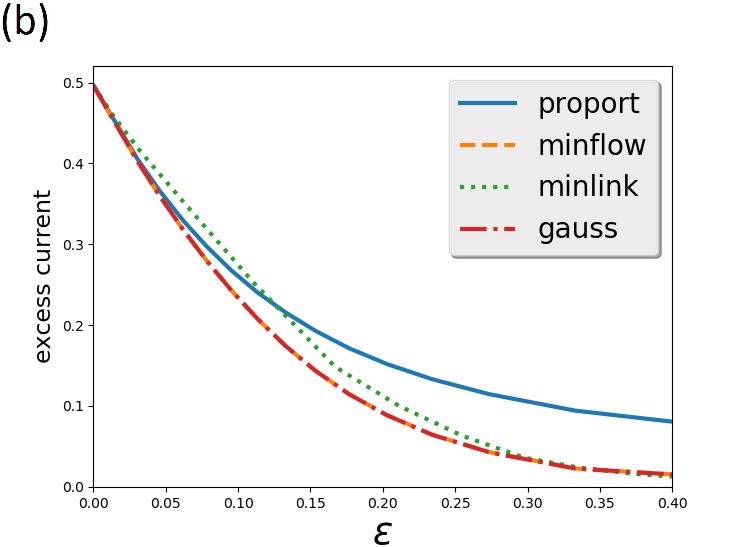}
\par\end{centering}
\caption{\label{fig:levy_stable}Simulation results of (a) the fraction of overloaded links and (b) the amount of excess current per link in units of $\sqrt{\langle\delta y_{ij}^2\rangle}$, for the bandwidth allocation schemes in the RTS96 network as a function of the tolerance factor $\varepsilon$ when the flow fluctuations follow the L\'evy-stable distribution respectively. {\it gauss} corresponds to bandwidth allocation scheme according to Eq. \eqref{band_flow}.}
\end{figure}

The {\it minflow} algorithm is given by
\begin{equation}
\Delta L_{ij} = S^{-1}(\lambda|c_{ij}),
\label{eq:Levy_minflow}
\end{equation}
\begin{equation}
S(x|c) = \int_x^\infty dy P(y|c).
\end{equation}
Since the L\'evy-stable distribution does not have an analytical expression in general, we obtain the optimal bandwidth allocation by solving Eq. \eqref{eq:Levy_minlink} and Eq. \eqref{eq:Levy_minflow} numerically. As a demonstration, we perform simulations in the RTS 96 network with the resource fluctuations following the L\'evy-stable distributions with scale parameter $c=1$. Figure \ref{fig:levy_stable} shows that both {\it minlink} and {\it minflow} algorithms are more robust and converge to zero much faster than the proportionate bandwidth allocation scheme. For comparison, we also plot the performances of the algorithms ignoring the long-tail nature of the resource distributions, that is, algorithms assuming that the resource distributions are Gaussian ones with their variances equal to the scaling parameters (i.e. Eq. \eqref{band_flow}). Figure \ref{fig:levy_stable}(a) shows that the Gaussian assumption leads to suboptimal results of the fraction of overloaded links, whereas Fig. \ref{fig:levy_stable}(b) shows that the Gaussian assumption results in effectively the same performance as the optimum.

\section{\label{sec:Optimal-Resource-Adjustment}Optimal Resource Adjustment} 

In the previous section, we study how to enhance the robustness of the network against fluctuations through increasing the total bandwidth. However, such an approach is usually more useful in network design. For real-time control, it is more practical to adjust the resources in nodes such as the implementation of load shedding in power engineering \citep{bienstock2011optimal}. In dynamic control, usually only short-term predictions of fluctuations of power supply and demand in the form of probabilistic distributions are available. Therefore in this section, we study how to optimally adjust the resources in the nodes to increase the stability of the network given the probabilistic information of the fluctuations.    

We begin by writing the resource $\Lambda_i$ as 
\begin{equation}
\Lambda_i = \Lambda_i^0 + \delta\Lambda_i + \pi_i,
\end{equation}
where $\Lambda_i^0$ is the original resource, $\delta\Lambda_i$ is the resource fluctuations and $\pi_i$ is the controllable component. For a power grid, one can treat the controllable component as the controllable power sources such as the power generated from natural gas or the load to be shed. When there are fluctuations in the resources, the induced flow fluctuations can cause overload in links and we consider link $(ij)$ to be overloaded when $|y_{ij}| > L_{ij}$ as in the previous section. For simplicity, we set $L_{ij} = |y_{ij}^0|$ in this section. With the introduction of $\pi_i$, the current flow $y_{ij}$ can be written as 
\begin{equation}
y_{ij} = y_{ij}^0 + \delta y_{ij} +z_{ij},
\end{equation}
where $z_{ij}=\sum_l (G_{jl} -G_{il})\pi_l$. Hence, with a proper value of $\pi_i$, the current flow can be reduced and our objective is to find the optimal $\pi_i$ such that the total number of overloaded links is minimized. Suppose the resource fluctuations follow the Gaussian distribution, then the flow fluctuations will also follow the Gaussian distribution and the objective function to be minimized is given by 
\begin{equation}
\begin{aligned}
F &=\frac{1}{E}\sum_{(ij)}\int_{L_{ij} - \text{sgn}(y_{ij}^0)(y_{ij}^0 + z_{ij})}^{\infty} P_{ij}(\delta y) d(\delta y) \\
&= \frac{1}{2E} \sum_{(ij)}\text{erfc} \left( \frac{\text{sgn}(-y_{ij}^0)z_{ij}}{\sqrt{2\langle\delta y_{ij}^2 \rangle}}\right)
\end{aligned}
\end{equation}
where $E$ is the total number of links. To have a practical resource adjustment, we introduce constraints to restrict the value of $\pi_i$. As there are no excess resources in the network, the sum of $\pi_i$ have to be equal to zero (i.e. $\sum_i \pi_i=0$). Moreover, to have a fair comparison between different adjustment schemes, we introduce an additional constraint to restrict the total amount of changes in the resources, $\sum_i |\pi_i| = c\sum_i |\Lambda_i^0|$, where $c$ is a parameter to control the total amount of changes. In fact, one can view $c$ as the mean ratio to be changed in the resource for each node. Furthermore, we also restrict $\pi_i$ to reduce the value of $\Lambda_i^0$ (i.e. $\rm{sgn}(\Lambda_i^0)\pi_i \leq 0$) and that the changes do not reverse the nature of supply and demand of the nodes (i.e. $-\rm{sgn}(\Lambda_i^0)(\pi_i+\Lambda_i^0) \leq 0$). Combining all the constraints, we propose an optimal resource adjustment scheme by solving the following constrained optimization problem,
\begin{equation*}
\begin{aligned}
& \underset{\{\pi_i\}}{\text{minimize}} 
& & \frac{1}{2E} \sum_{(ij)} \text{erfc} \left( \frac{\text{sgn}(-y_{ij}^0)z_{ij}}{\sqrt{2\langle\delta y_{ij}^2 \rangle}} \right) \;, \\
& \text{subject to}
& & \sum_i |\pi_i| = c\sum_i |\Lambda_i^0| \;, \\
&&& \sum_i\pi_i = 0 \;, \\
&&& \text{sgn}(\Lambda_i^0)\pi_i \leq 0 \;, \\
&&& -\text{sgn}(\Lambda_i^0)(\Lambda_i^0 + \pi_i) \leq 0 \;.
\end{aligned}
\end{equation*}
From the form of the optimization problem, one can see that the proposed optimal resource adjustment scheme has some similarities in the constraints to the standard linear optimal power flow (OPF) control scheme in electricity systems. Yet, the objective of OPF is to set the generator outputs to meet the demand at the  minimum cost while the proposed optimal resource adjustment scheme reduces both the supply and demand of the resources such that the average expected overload probability of the links can be minimized.

We have also attempted an alternative formulation of the resource adjustment scheme by minimizing the total load to be shed, subject to an upper bound of the overload probability. In other words, the role of the individual terms in the objective function and the first constraint are interchanged in this formulation. Then it turns out that the overload probability can be inverted to give a linear constraint in terms of $z_{ij}$, and the optimization problem reduces to a linear programming problem. The solution to this linear programming problem resides in the vertex of the hyperpolygon of the solution space with the resource of some nodes completely turned off. This reveals that the approach is incomplete, and additional constraints beyond the scope of this study have to be added. Therefore we will not pursue this approach here.

\begin{figure}
\begin{centering}
\includegraphics[scale=0.5]{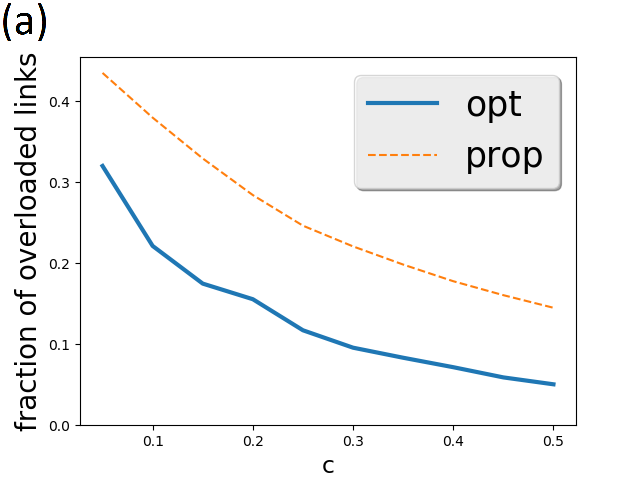}
\includegraphics[scale=0.5]{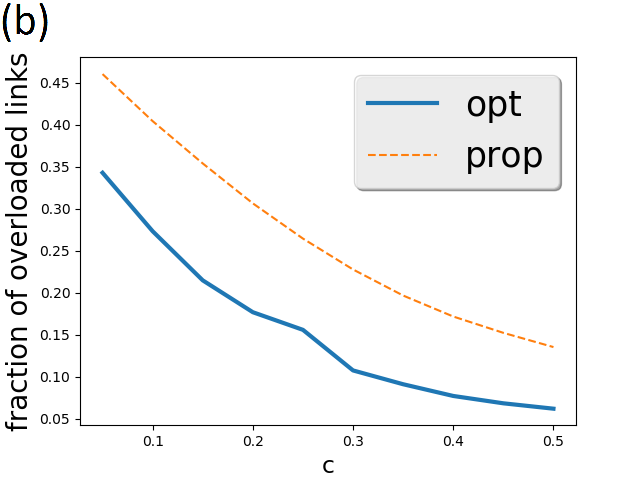}
\par\end{centering}
\caption{\label{fig:rts_load_shed}Simulation results of the fraction of overloaded links in (a) the RTS96 network and (b) IEEE300-Bus network averaged over 100 samples as a function of the load-shedding ratio $c$. For simplicity, {\it proport} and {\it opt} corresponds to the proportionate resource adjustment scheme and optimal resource adjustment scheme respectively.}
\end{figure}

Since there is no analytical solution for the above optimization problem, one has to solve it numerically. Since the equality and inequality constraints are linear in the control variables $\pi_i$, it can be solved using the barrier methods \citep{boyd2004convex}. For comparison in the RTS 96 network and the IEEE300-Bus network, we introduce the proportionate reduction scheme $\pi_i^{prop} = -c(\Lambda_i^0)$ which proportionately reduces the current flow in the network. $\Lambda_i^0$ follows a Gaussian distribution with mean equal to 0 and variance equal to 1 with the fluctuation $\delta\Lambda_i$ following a Gaussian distribution with mean equal to 0 and variance equal to $0.1(\Lambda_i^0)^2$ independently. Figure \ref{fig:rts_load_shed} shows that in the optimal adjustment scheme, the network has a much higher stability compared with the proportionate reduction.

\begin{figure}
\begin{centering}
\includegraphics[scale=0.65]{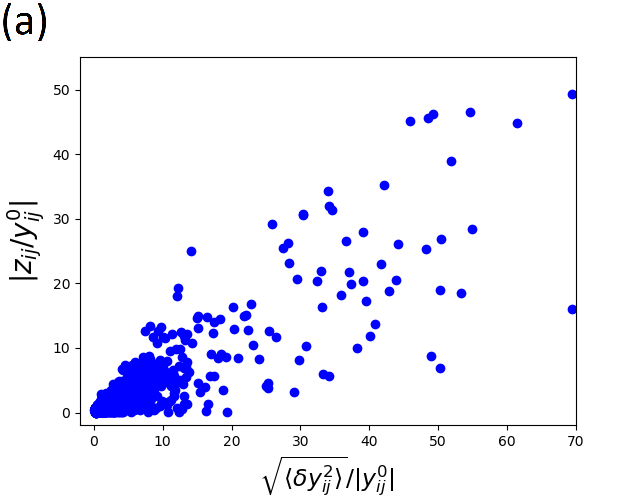}
\includegraphics[scale=0.45]{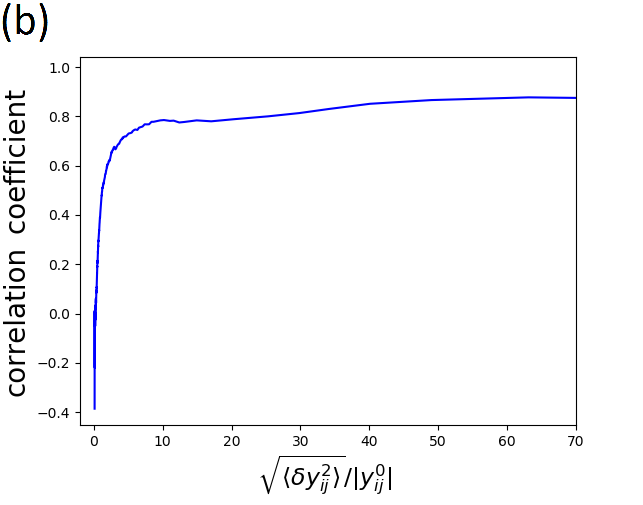}
\par\end{centering}
\caption{\label{fig:compare_ls_band}Comparison of the link behaviors in the optimal bandwidth allocation and resource adjustment schemes. We have used 30 samples of  ER network with $N=100$ and $p=0.05$. The resources and resource fluctuations have the same setting as in Fig. \ref{fig:rts_load_shed}. (a) $|z_{ij}/y_{ij}^0|$ versus $\sqrt{\langle\delta y_{ij}^2 \rangle}/|y_{ij}^0|$ with $c=0.3$. (b) Pearson correlation coefficient versus $\sqrt{\langle\delta y_{ij}^2 \rangle}/|y_{ij}^0|$.}
\end{figure}

Similar to the optimal bandwidth allocation scheme, the current reduction of the links in the optimal resource adjustment scheme is correlated with the strength of the flow fluctuations. In fact, links that are given larger bandwidths in the optimal bandwidth allocation scheme will usually experience a larger reduction in current flow according to the optimal resource adjustment scheme. From Fig. \ref{fig:compare_ls_band}(a), one can see that those links having high current reduction in the resource adjustment scheme usually have a larger bandwidth allocated according to Eq. \eqref{band_flow}. For each value of rescaled fluctuation $\sqrt{\langle\delta y_{ij}^2 \rangle}/|y_{ij}^0|$, we compute the Pearson correlation coefficient between $|z_{ij}/y_{ij}^0|$ and $\sqrt{\langle\delta y_{ij}^2 \rangle}/|y_{ij}^0|$ in the range spanned from its value down to 0. Figure 6(b) shows that the correlation is positive.

\begin{figure}
\begin{centering}
\includegraphics[scale=0.6]{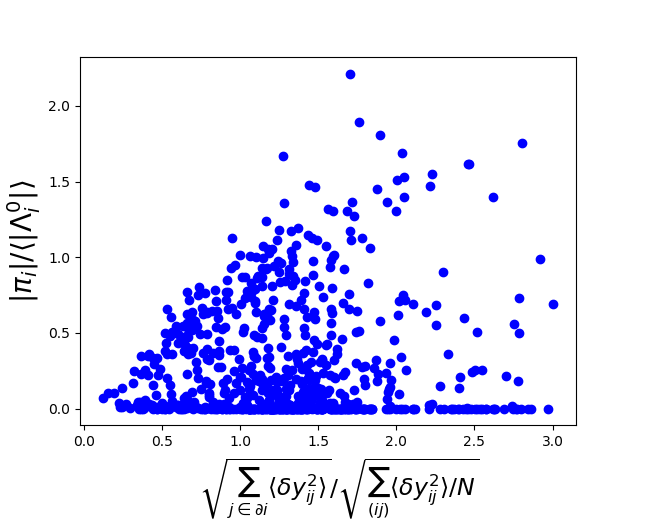}
\par\end{centering}
\caption{\label{fig:opt_change}Comparison of the node behaviors in the optimal resource adjustment scheme (in terms of the rescaled resource adjustment $|\pi_i|/\langle\Lambda_i^0|\rangle$) and the optimal bandwidth allocation scheme (in terms of the rescaled bandwidth adjustment $\sqrt{\sum_{j\in\partial i}\langle\delta y_{ij}^2\rangle}/\sqrt{\sum_{(jk)}\langle\delta y_{jk}^2\rangle/N}$). The network and parameter setting are the same as Fig. \ref{fig:rts_load_shed} with $c=0.3$ (30 samples).}
\end{figure}

\begin{figure}
\begin{centering}
\includegraphics[scale=0.4]{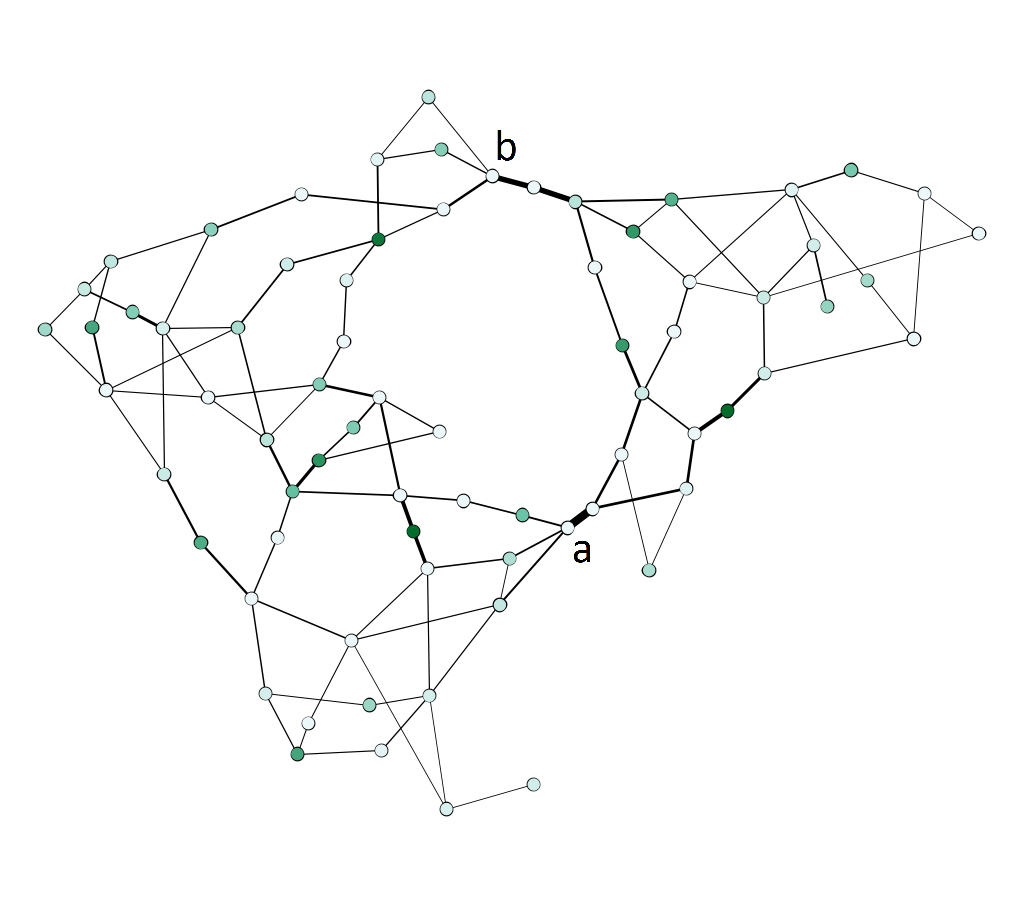}
\par\end{centering}
\caption{\label{fig:rts}Illustration of the relay effect in the RTS 96 network with the same parameter setting as Fig. \ref{fig:rts_load_shed} [as mentioned in the text]. The darker the color of the node, the larger the changes $|\pi_i|$. The thickness of the edge represents the strength of the flow fluctuations $\langle \delta y_{ij}^2\rangle$.}
\end{figure}

We further compare the behaviors of the nodes under the optimal resource adjustment and bandwidth allocation schemes. Figure \ref{fig:opt_change} shows the resource adjustment $|\pi_i|$ against the total flow fluctuation $\sqrt{\sum_{j\in\partial i} \langle \delta y_{ij}^2 \rangle}$ through node $i$ in the RTS 96 network. From the figure, one can see that most nodes connected with links having large flow fluctuations require a larger reduction in resources. However, there are also considerable number of nodes having large flow fluctuations that do not have any changes in resources $\pi_i=0$. In fact, those nodes can be viewed as nodes used for relaying the 'reduction' of current flow $z_{ij}$. Since those nodes are connected with links having large flow fluctuations, their connected links require a larger reduction in current flow such that the overload probability can be minimized. To satisfy the constraints in the optimization problem, some nodes are therefore used for relaying the flow reduction without themselves participating in load reduction. Usually, those nodes are connected with nodes that have a large reduction in resources. As an illustration, Fig. \ref{fig:rts} shows the RTS 96 network with the same parameter setting as Fig. \ref{fig:rts_load_shed}. In this network, nodes a and b have no change in resources during load shedding at $c = 0.3$, but they play the role of relaying resources to neighboring nodes undergoing large resource reductions. 

\section{Conclusion} 
We have formulated the chemical potential method for finding the current flows in the transportation network with a general cost function. The method can be applied to optimize stability in transportation networks and also power grids in the DC approximation. With the discrete Green's functions, one can estimate the mean and variance of the flow fluctuations induced by the resource fluctuations. 

Based on the precise variance estimates, we have developed the optimal bandwidth allocation scheme that can minimize the fraction of overloaded links or the total excess currents in networks with fluctuations. In contrast, we have shown that using the traditional proportionate bandwidth allocation cannot effectively increase the robustness of the networks against fluctuations. Analytical results assuming that the link fluctuations are independent have excellent agreement with simulation results. Both results show that the optimized methods of allocating bandwidth can effectively enhance the stability of the networks against fluctuations compared with the proportionate bandwidth allocation, reducing the asymptotic loss from a power law to one with a Gaussian tail. Consequently, the scheme minimizes the bandwidth resources dramatically while achieving the same level of stability. 

The optimization scheme is based on the precise estimate of resource fluctuations. While we have demonstrated its effectiveness assuming Gaussian fluctuations, we have also verified its applicability to long-tailed distributions such as the L\'evy-stable distribution.

Moreover, we have developed a scheme to optimally adjust the resources in the network such that it can minimize the probability of link overload in the presence of fluctuations.. Simulation results show that using the proposed resource adjustment scheme can have a better performance than the proportionate resource reduction. Furthermore, we found that there is a close correlation between the optimal bandwidth allocation and resource adjustment schemes. The former assigns more bandwidths to links with large flow fluctuations while the latter reduces more current flow along them. Exceptions are some nodes connected to highly fluctuating links that relay their fluctuations to other links without their own participation in resource adjustments. 

The scheme can be generalized to optimizing other cost functions. An example is the synchronizability in AC networks \citep{li2017optimizing}. We have studied the enhancement of synchronizability in networks with resource fluctuations, but due to the different setting, the variance estimates are less important, because we found that by optimizing the mean value of the synchronizability, the variance is also reduced \citep{tsang2018enhancing}.

Another possible extension of our work is to investigate the bandwidth allocation and resource adjustment for power transmission among different countries and regions of the continental power grid. For such applications, the optimization will be based on the total resource and total resource fluctuations of each region, and their scaling with regional size will be relevant \citep{schafer2017scaling}.

Besides power grids, many other network problems can be formulated as flow optimization problems. They also have wide applications in transportation networks and communication networks \citep{23sparse_graph, shao2007optimal, wong2006equilibration}. The network flow optimization problems are usually approached by finding a specific cost function to minimize. Hence the techniques developed in this paper can be extended to other network flow problems with adapted cost functions.

Another approach to estimate the mean and variance of network currents is the message-passing algorithm, also known as belief propagation or the cavity method \citep{mezard1987spin, pearl2014probabilistic}. For discrete resource variables, the message-passing algorithm is only approximate in loopy networks, but for real-valued resource variables, the means of the resource variables estimated by the algorithm are known to be exact even when loops are present in the network structure if it converges, and convergence is prevalent in most cases \citep{23sparse_graph, weiss2000correctness}. Since the algorithm involves local updates only, it is advantageous in distributed control, especially for large networks or evolving networks. However, when variances of the currents are calculated, the message-passing under-estimates the variances due to it negligence of loop effects. Nevertheless, for known current variances, the algorithm has been formulated to calculate the optimal current in volatile networks \citep{harrison2016optimal}. 

The optimal adjustment scheme developed have only consider one-step control. In general, the power to be ramped down from the generators should be tuned gently to their current output as generators cannot modify their output arbitrarily fast. It was also found that applying load shedding at an early stage may not be optimal \citep{bernstein2014power}. Hence, using the technique developed, we can extend the load shedding problem to a scheduling problem through multiple-step control.

While our study is motivated by problems of power grid stability such as  cascading failures, especially in the modern era of renewable energy and power trading, power grid stability in reality involves factors far more complex than the simplistic model considered in this work. Such factors include the short-time dynamics, correlations in wind and solar power generation, heterogeneous distribution of load and generation, and technical properties of different lines on different voltage levels of the power grid. Nevertheless, it is hoped that the approaches introduced herein are able to inspire different approaches in more realistic models, and further development will benefit from inter-disciplinary collaborations.


\begin{acknowledgments}
We thank Juntao Wang and He Wang for fruitful discussions. This work is supported by grants from the Research Grants Council of Hong Kong (grant numbers 16322616, 16306817 and 16302419).
\end{acknowledgments}


\appendix

\section*{\label{sec:appendix}Appendix: PERFORMANCES of {\it Proport}, {\it Minlink} and {\it Minflow}}

\subsection{The Proportionate Scheme}

Consider the resource fluctuations follow Gaussian distribution with mean equal to 0 and $\langle\delta\Lambda_i^2\rangle = k^2(\Lambda_i^0)^2$, where $k$ is the fluctuation parameter. Then the flow fluctuations will also follow a Gaussian distribution with mean 0 and variance
\begin{equation}
\sigma_{\delta y}^2 = k^2\sigma_y^2.
\end{equation}
To study the performance of the bandwidth allocation schemes, we express the overload probability and the excess current in a link $\left( ij\right)$ as a function of $\Delta L_{ij}$. The overload probability of a link $\left(ij\right)$ is given by
\begin{equation}
P_O^{ ( ij ) } = \int_{\Delta L_{ij}/\sigma_{\delta y}}^{\infty} Dz = H\left(\frac{\Delta L_{ij}}{\sigma_{\delta y}}\right) ,
\end{equation}
where $Dz \equiv G(z) dz$ with $G(z) \equiv \exp(-z^2/2)/\sqrt{2\pi}$  and $H(x)\equiv\int_x^\infty Dz$. For the excess current in a link $(ij)$, we have
\begin{equation}
P_C^{ ( ij ) } = \int_{\Delta L_{ij}/\sigma_{\delta y}}^{\infty} Dz (z\sigma_{\delta y} - \Delta L_{ij}).
\end{equation}

Consider the behavior of the proportionate bandwidth allocation (i.e. $\Delta L_{ij} = \varepsilon |y_{ij}|$). The average overload probability is given by
\begin{equation}
P_O^{prop} = 4\int_{0}^{\infty}D y_0 \int_{0}^{\infty} Dy H \left(\frac{\varepsilon\sigma_y y_0}{\sigma_{\delta y} y} \right).
\end{equation} 
Interchanging the order of integration of $y_0$ and $y$, 
\begin{equation}
P_O^{prop} = \frac{2}{\pi} \int_{0}^{\infty}Dy \tan^{-1}\left(\frac{k y}{\varepsilon}\right).
\end{equation}
For $\varepsilon \gg 1$, we have
\begin{equation}
\label{app_prop_fail}
P_O^{prop} \approx \frac{2k}{\pi\varepsilon} \int_{0}^{\infty} Dy y = \sqrt{\frac{2}{\pi^3}}\frac{k}{\varepsilon}.
\end{equation}
For the average excess current using the proportionate bandwidth allocation scheme, we obtain 
\begin{equation}
P_C^{prop} = 4\int_{0}^{\infty}Dy_0 \int_{0}^{\infty} Dy \int_{\varepsilon y_0 / k y}^{\infty} Dz(\sigma_{\delta y} yz - \varepsilon\sigma_y y_0).
\end{equation}
To simplify, we interchange the order of the integration
\begin{equation}
P_C^{prop} = 4 \sigma_{\delta y} \int_{0}^{\infty}Dy y \int_{0}^{\infty} Dy_0 \int_{\varepsilon y_0 / k y}^{\infty} Dz\left(z - \frac{\varepsilon y_0}{k y}\right).
\end{equation}
By transforming to polar coordinates in the space of $y_0$ and $z$, one can obtain
\begin{equation}
P_C^{prop} = \sigma_{\delta y} \sqrt{\frac{2}{\pi}} \int_{0}^{\infty} Dy \frac{k y^2}{\sqrt{k^2 y^2 +\varepsilon^2} +\varepsilon}.
\end{equation}
In the limit of $\varepsilon \gg 1$, the above expression can be simplified as 
\begin{equation}
\label{app_prop_current}
P_C^{prop} = \frac{k\sigma_{\delta y}}{2\varepsilon}\sqrt{\frac{2}{\pi}}\int_{0}^{\infty}Dy y^2 = \frac{k\sigma_{\delta y}}{\sqrt{8\pi}\varepsilon}.
\end{equation}

\subsection{The Minlink Scheme}

We further study the behavior of the {\it minlink} algorithm in which $\Delta L_{ij}$ is given by Eq. \eqref{link_band_cond}. Let $\rho = 1/(\lambda\sqrt{2\pi}\sigma_{\delta y})$. From the constraint Eq. \eqref{constraint_link}, we have
\begin{equation}
2\int_0^{\rho} Dy \sigma_{\delta y} y\sqrt{2\ln\left(\frac{\rho}{y}\right)} = 2\varepsilon\int_{0}^{\infty}Dz\sigma_y z.
\end{equation}
This results in an equation for $\varepsilon$,
\begin{equation}
	\varepsilon = k\sqrt{2\pi}
	\int_0^\rho Dy y
	\sqrt{2 \ln\left(	\frac{\rho}{y} \right)}.
\label{eq:minlink_eps}
\end{equation}
The average overload probability using the {\it minlink} algorithm is given by
\begin{equation}
	P_O^{link} = H(\rho)
	+2\int_0^\rho Dy
	H\left(\sqrt{2 \ln\left( \frac{\rho}{y} \right)} \right).
\label{eq:minlink_PO}
\end{equation}
In the limit $\varepsilon \gg 1$, Eq. \eqref{eq:minlink_eps} reduces to
\begin{equation}
	\rho\approx\exp\left(
	\frac{\varepsilon^2}{2k^2}
	\right),
\end{equation}
and the sum in Eq. \eqref{eq:minlink_PO} is dominated by by the second term. Using the approximation $H(x)\approx\exp(-x^2/2)/(\sqrt{2\pi}x)$ when $x\gg 1$, and neglecting logarithmic terms, we have
\begin{equation}
P_O^{link} \approx \frac{k}{\pi\varepsilon}\exp\left(-\frac{\varepsilon^2}{2 k^2}\right).
\label{app_minlink_fail}
\end{equation}

Similarly, the average excess current is given by
\begin{equation}
\begin{aligned}
	P_C^{link} = &2\int_\rho^\infty Dy \int_0^\infty Dz 
	\sigma_{\delta y} y z \\
	&+ 2\int_0^\rho Dy 
	\int_{\sqrt{2\ln(\rho/y})}^\infty Dz
	\sigma_{\delta y} y
	\left( z - \sqrt{2\ln(\rho/y)}\right),
\end{aligned}
\end{equation}
which simplifies to
\begin{equation}
	P_C^{link} = \sqrt{\frac{2}{\pi}}G(\rho)
	+ 2\int_0^\rho Dy y
	[G(z) - z H(z)]_{z=\sqrt{2\ln(\rho/y)}}.
\end{equation}
In the limit $\varepsilon \gg 1$, we have
\begin{equation}
	P_C^{link} \approx
	\frac{\sigma_{\delta y}k}{\pi\varepsilon}
	\exp\left(-\frac{\varepsilon^2}{2k^2}\right).
\label{app_minlink_current}
\end{equation}

\subsection{The Minflow Scheme}

Finally, for the {\it minflow} algorithm, we can express $\varepsilon'$ in terms of $\varepsilon$ using Eq. \eqref{constraint_link},
\begin{equation}
	\varepsilon'\sum_{(ij)}\sqrt{\langle\delta y_{ij}^2\rangle}
	=\varepsilon\sum_{(ij}|y_{ij}|.
\end{equation}
Using Eqs. \eqref{var_ychemsim} and \eqref{current_chem}, we can write
\begin{equation}
	\varepsilon' k \sum_{(ij)}\sqrt{\sum_l
	(G_{jl} - G_{il})^2 \Lambda_l^2}
	=\varepsilon\sum_{(ij)}|\sum_l
	(G_{jl} - G_{il})\Lambda_l|.
\end{equation}
Note that inside the square root sign on the left hand side, the term is the square of a Gaussian variable with mean 0 and variance $N \langle (G_{jk} - G_{ik})^2\rangle \langle\Lambda_k^2\rangle$. The term inside the absolute value sign on the right hand side is also a Gaussian variable with mean 0 and variance $N \langle (G_{jk} - G_{ik})^2\rangle \langle\Lambda_k^2\rangle$. Hence we have
\begin{equation}
	\varepsilon' k N\int Dy y^2
	=\varepsilon N\int Dy |y|,
\end{equation}
implying that
\begin{equation}
	\varepsilon' = \sqrt{\frac{\pi}{2}}\frac{\varepsilon}{k}.
\end{equation}
The average overload probability using {\it minflow} is given by
\begin{equation}
\label{app_minflow_fail}
P_O^{flow} = 2\int_{0}^{\infty} DyH\left(\frac{\varepsilon'\sigma_{\delta y}y}{\sigma_{\delta y}y}\right) = H(\varepsilon'),
\end{equation}
and the average excess current is given by
\begin{equation}
\label{app_minflow_current}
\begin{aligned}
P_C^{flow} &= 2\int_{0}^{\infty}Dy \int_{\varepsilon'}^{\infty}Dz(\sigma_{\delta y} yz - \varepsilon'\sigma_{\delta y} y) \\
	&=\sqrt{\frac{2}{\pi}}\sigma_{\delta y}
	[G(\varepsilon') - \varepsilon' H(\varepsilon')].
\end{aligned}
\end{equation}
In the limit $\varepsilon \gg 1$, we obtain
\begin{equation}
	P_O^{flow} \approx
	\frac{k}{\sqrt{2\pi}\varepsilon}
	\exp\left(
	-\frac{\varepsilon^2}{2k^2}\right),
\end{equation}
\begin{equation}
	P_C^{flow} \approx
	\frac{\sigma_{\delta y}k}{\pi\varepsilon}
	\exp\left(
	-\frac{\varepsilon^2}{2k^2}\right).
\end{equation}

The proportionate allocation scheme yields a fraction of overloaded links and an average excess current scaling as $\varepsilon^{-1}$ in the limit of large $\varepsilon$ as shown in Eq. \eqref{app_prop_fail} and \eqref{app_prop_current} respectively. This behavior of convergence to zero is much weaker than the Gaussian tail convergence in the $minlink$ (Eq. \eqref{app_minlink_fail} and \eqref{app_minlink_current}) and $minflow$ (Eq. \eqref{app_minflow_fail} and \eqref{app_minflow_current}) schemes.


\end{document}